\documentclass[prd,showpacs,preprintnumbers,nofootinbib,aps,twocolumn]{revtex4}
\usepackage{subeqnarray}
\usepackage{epsfig,amsmath,amssymb}
\usepackage{mathrsfs}
\usepackage{enumerate}
\usepackage[usenames,dvipsnames]{color}
\usepackage[pagebackref=true, colorlinks=true]{hyperref}
\definecolor{redish}{rgb}{0.7,0.2,0.0}  
\definecolor{bluish}{rgb}{0.2,0.5,0.8}
\hypersetup{linkcolor=redish,          
                  citecolor=blue,        
                  filecolor=magenta,      
                  urlcolor=bluish}          

\DeclareFontFamily{U}{rsfs}{}         
\DeclareFontShape{U}{rsfs}{m}{n}{<5> rsfs5 <6><7> rsfs7          %
  <8><9><10><10.95><12><14.4><17.28><20.74><24.88> rsfs10}{}     %
\DeclareMathAlphabet{\mathfs}{U}{rsfs}{m}{n}                     %
\newcommand{\mfs}[1]{\mathfs {#1}}                              %

\newcommand{\ba}{\nopagebreak[3]\begin{eqnarray}}
\newcommand{\ea}{\end{eqnarray}}
\newcommand{\bii}{\begin{itemize}}
\newcommand{\eii}{\end{itemize}}
\newcommand{\nn}{\nonumber}

\newcommand{\sO}{{\mfs O}}

\newcommand{\f}{\frac}

\def \d{\delta}

\def \g{\gamma}

\def \j{\sqrt{j(j+1)}}

\def \lm{\lambda}

\def \s{\sigma}

\def \sj{s_j^{\star}}

\def \O{\Omega}

\def \({\left(}
\def \){\right)}
\def \[{\left[}
\def \]{\right]}
\begin{document}
\title{An area rescaling ansatz and black hole entropy from loop quantum gravity}
\author{Abhishek Majhi}%
\email{abhishek.majhi@gmail.com}
\affiliation{Instituto de Ciencias Nucleares\\
Universidad Nacional Aut\'onoma de M\'exico\\
A. Postal 70-543, Mexico D.F. 04510, Mexico}

\begin{abstract}
Considering the possibility of `renormalization' of the gravitational constant on the horizon,  leading to a dependence on the level of the associated Chern-Simons theory, a rescaled area spectrum is proposed for the non-rotating black hole horizon in loop quantum gravity.  The statistical mechanical calculation leading to the entropy provides a unique choice of the rescaling function for which the Bekenstein-Hawking area law is yielded without the need to choose the Barbero-Immirzi parameter $(\g)$. $\g$ is determined by studying the limit in which  the `renormalized' gravitational constant on the horizon asymptotically approaches the `bare' value.  Unlike the usual, much criticized, practice of choosing $\g$ just for the sake of the entropy matching the area law, its value is now rather determined by a physical consistency requirement. 
\end{abstract}
\maketitle
\section{Introduction}
Loop quantum gravity (LQG) provides a platform for the calculation of  entropy for non-rotating\footnote{The adjective `non-rotating' will be dropped and assumed, henceforth.} black holes from the first principles, albeit in the kinematic framework \cite{qg2}. The main criticism of this approach has been the necessity to choose a particular value of the Barbero-Immirzi parameter $(\g)$, which is a dimensionless constant that characterizes the family of inequivalent kinematic quantization sectors of LQG, to obtain the Bekenstein-Hawking area law (BHAL) \cite{jac}. If the derivation is correct, then it is expected that one should get the BHAL without having to choose $\g$. As it appears, the full knowledge of the dynamics of LQG, the horizon degrees of freedom and the correct semi-classical limit of the theory are required to achieve this goal \cite{jac}, which, unfortunately, does not seem to become available in near future. Nonetheless, the kinematic framework holds the potential to give us the hints towards the correct physical elements that give rise to the black hole entropy, which, in turn may lead the path towards understanding the underlying dynamics. Here, I shall point out that there is a possibility of the involvement of a `renormalization'\footnote{Since the arguments are based on analogy, the words associated with renormalization are kept in quotes.} of the gravitational constant {\it on the horizon} and incorporation of this effect in the entropy calculation  leads us to the BHAL from LQG {\it without having to choose $\g$}. I shall heuristically argue that the quantum field theoretic structure that effectively describes the horizon degrees of freedom, suggests that there is a possibility for a {\it rescaled} area spectrum to be used for the black hole horizon in LQG due to the `renormalization' of the gravitational constant {on the horizon}. Further, the calculation of entropy with this rescaled area spectrum provides us with the unique  rescaling function that leads to the BHAL without having to choose $\g$. The value of $\g$ is {\it determined}, irrespective of obtaining the BHAL, by studying how the `renormalized' gravitational constant on the horizon should {\it asymptotically} approach its `bare' value in a limit that can be viewed as the `fixed point' of the `renormalization group flow' on the horizon. 
The novelty of this, albeit heuristic, work lies in the fact that the value of $\gamma$ is now determined by a physical consistency requirement rather than being chosen just to match a desired result.



\section{Motivation}
The entire procedure of the entropy calculation for black holes in LQG consists of the following steps:

\vspace{0.2cm}
1) {\it Horizon field dynamics}: The effective quantum field dynamics on the horizon (of topology $S^2\times R$) is governed by a quantum Chern-Simons (CS) theory on a punctured 2-sphere and these punctures act as point-like sources coupled to the CS field strength \cite{qg2}. The Hilbert space of this theory provides the state space of the horizon degrees of freedom that give rise to the entropy \cite{qg2,wit,km98}.

\vspace{0.2cm}
2) {\it Spectrum of the source}: Consider any arbitrary geometric 2-surface that is topologically $S^2$. The quantum area of such a surface, in LQG, is given by $8\pi\g G\sum_{l=1}^N\sqrt{j_l(j_l+1)}$ (setting $\hbar=c=1$) and any $j$ can take values like $0,\f{1}{2},1,\cdots, \infty$. $j_1,j_2,\cdots, j_N$ are the quantum numbers carried by the intersection points (punctures) of the spin network edges with that 2-surface, $N$ being the total number of punctures \cite{area}. This is the same area spectrum that is used during the entropy calculation for black holes, with a crucial modification due to the interplay between quantum geometry and the CS theory on the horizon,  that $j$ can take values like $\f{1}{2},1,\cdots, \f{k}{2}$, where $k:=A_c/4\pi\g G$, $A_c$ being the classical area of the black hole \cite{qg2}. Hence, the contribution from an individual puncture (point-like source of the CS theory on the horizon) is $8\pi\g G\sqrt{j(j+1)}$ with $j\in\{\f{1}{2},1,\f{3}{2},\cdots,\f{k}{2}\}$.

\vspace{0.2cm}

3) {\it Statistical mechanics}: Having the estimate of the microstate count from the first step and the area spectrum of the black hole from LQG in the second step, the statistical mechanics is applied to calculate the entropy.


Now, let me focus on the second step. It implies, in principle, the quantum area of an arbitrary geometric 2-surface of topology $S^2$, can be infinite, irrespective of the classical area of the surface. So, it is expected on physical grounds that this should {\it not} be the case when the concerned 2-surface is that of a physical object and the value of $j$ should acquire an upper cut-off provided by the underlying physics associated with the surface of the physical object. This is exactly what happens for the black hole horizon. The value of  $j$ acquires an upper bound $k/2$, where the $k$ is the level of the CS theory associated with the horizon i.e. the first step plays a crucial role. Therefore, the theory governing the physics associated with the horizon naturally provides this upper bound. This is a result which is already manifest from the LQG kinematics and the effective horizon theory. However, the lack of knowledge about the full dynamics of a quantum black hole in LQG leaves room for some physics,  associated with the horizon degrees of freedom contributing to the entropy of a black hole, that may be missing in the kinematics. As I shall argue,  the information that is already available from the kinematics (the first step), indeed,  hints towards such a possibility. 

  The field equations on a black hole horizon is that of a CS theory coupled to sources:
\ba
F=\f{1}{k}\cdot\Sigma\label{bc}
\ea
where $F$ is the curvature of the CS gauge fields on the horizon and $\Sigma$ are the sources from the bulk.  In the quantum theory, the source $\Sigma$ is non-zero only at the punctures. Effectively, the theory on the horizon is a quantum CS theory coupled to point-like sources on a 2-sphere. The spectrum  associated with a single source is calculated for an arbitrary 2-sphere where there is no coupling with any field strength and it is given by
\ba
a_j=8\pi\g G\j.\label{ab}
\ea
So, these punctures on an arbitrary 2-sphere are like `free' excitations and the spectrum in eq.(\ref{ab}) can be regarded as `bare' spectrum. However, these `free' excitations get coupled to the CS field strength in case the 2-sphere is a cross-section of a black hole horizon. 

Now, in quantum field theory (QFT), the physical parameters like mass, charge, etc. associated with free particles get renormalized due to their coupling with fields, consequently affecting the mode spectrum. Analogously, in the present scenario, there is a possibility of $\g G$ in eq.(\ref{ab})\footnote{Since $\gamma$ and $G$ always appear as a product in the kinematics of LQG, one should consider the `renormalization' of $\g G$ rather than $G$ alone \cite{ben}.}, which can be viewed as the ``mode spectrum'' for the sources \cite{amindis},  getting `renormalized' due to the coupling with the CS field strength. This `renormalized' $\g G$, say $\tilde G$, should depend on  $k$, which is the cut-off for the allowed values of $2j$ that appears  naturally in the theory on the horizon resulting from its  gauge invariance \cite{qg2}. Since  the physical process involved with this `renormalization' is associated with the quantum theory {\it on the horizon} this $\tilde G$ can only affect the microscopic physics localized on the horizon.

  Although, this heuristic `renormalization' argument is only at the level of an analogy  made from a QFT viewpoint, nevertheless, the possibility of the scenario can not be completely ruled out unless one gets to know the full dynamics of the theory. 

\section{Rescaled area spectrum: an ansatz}\label{sec3}

As I have just argued, $\tilde G$ (the `renormalized' $\g G$), which enters the area spectrum of the black hole horizon, can only depend on $k$ and on the value of $\g$ because there are no other quantities intrinsic to the theory on the horizon. If $\delta G$ is the change in the value of $\g G$, then
\ba
\tilde G(k,\g)=\g G+\delta G(k,\g)
\ea 
Since $k$ and $\g$ are both dimensionless, simply on dimensional grounds, $\delta G\propto G$. Further, as $k\to\infty$, $\delta G$ must tend to zero because the sources get more weakly coupled to the CS field strength and the `renormalized' spectrum should approach towards the `bare' spectrum (this argument will be discussed more elaborately later). Hence, the most general form $\delta G$ can have is $\delta G(k,\g)= G\sum_{n\geq 1}\f{c_n}{k^n}$, where $c_n$ are some numbers that may involve $\g$.

Based on these arguments I propose that the area contribution from a single puncture with quantum number $j$, for a black hole horizon in LQG, be given by
\ba
a_j=8\pi\tilde G(k,\g)\j
\ea
where 
\ba
\tilde G(k,\g)=\s(k,\g)\g G\label{G}
\ea
is the `renormalized' gravitational constant on the horizon and $\s(k,\g)=1+\g^{-1}\sum_{n\geq 1}\f{c_n}{k^n}$. Hence, the spectrum of a source gets {\it rescaled}. In a nutshell, the expected properties of $\s$ are as follows:
\begin{enumerate}[(i)]
\item Since the `bare' spectrum needs to be positive definite, one has $\g>0$. It is needed that $\s>0$ so that the `renormalized' spectrum be positive definite too.
\item $\lim_{k\to\infty}\s(k,\g)=1$ i.e. as the sources get more weakly coupled to the CS field strength, $\tilde G$ asymptotically approaches $\g G$.
\end{enumerate}
As I shall show, the statistical mechanical calculation provides a unique choice of the function $\s$ that leads to the BHAL and satisfies property (i). Satisfaction of property (ii) by $\s$, which is a physical consistency requirement, will determine $\g$. It is crucial to note that property (i) and property (ii) are independent of each other.

\section{Entropy}
I shall consider here black holes with classical area $A_c (\gg\sO(G))$.
Quantum area of a cross-section of a black hole horizon, with spin configuration $\{s_j\}$:
\ba
A_q=8\pi\g \s G\sum_{j=1/2}^{k/2}s_j\j\label{reas}
\ea
where $k:=A_c/4\pi\g G$ and $s_j=$ number of punctures with quantum number $j$. Since $j$ ranges from $1/2$ to $k/2$, hence $k\geq 1 \implies \g\leq A_c/4\pi G$. Also, since $k$ is positive definite, the definition of $k$ suggests that $\g>0$. So, the quantum theory of the horizon is valid for $0<\g\leq A_c/4\pi G$.
Now, I shall implement the method of most probable distribution \cite{pathria} to calculate the microcanonical entropy of the black hole in the area ensemble. One can find the calculation (but, with the `bare' spectrum) in \cite{mpla1}. So, I shall provide the main steps and results here to avoid an unnecessary repeat.

The microstate count for a spin configuration $\{s_j\}$ for which $A_q=A_c\pm\sO(G)$:
\ba
\O[\{s_j\}]={N!}\prod_{j=1/2}^{k/2}\f{(2j+1)^{s_j}}{s_j!}
\ea
where $N:=\sum_{j=1/2}^{k/2}s_j$ and $\{s_j\}$ satisfies the following constraint (considering $A_c\gg\sO(G)$):
\ba
C: \sum_{j=1/2}^{k/2}s_j\j-A_c/8\pi\g \s G=0
\ea
Then one finds the most probable configuration (MPC) by solving the following equation:
\ba
\d \log\O[\{s_j\}]-\lm~\d C=0
\ea
where $\lm$ is the Lagrange multiplier. This yields the distribution for the MPC to be
\ba
\sj=N_0(2j+1)e^{-\lm\j}\label{mpc}
\ea
where $N_0:=\sum_{j=1/2}^{k/2}\sj$ and the entropy comes out to be
\ba
S=\(\f{\lm }{2\pi\g \s}\)\cdot\f{A_c}{4G}\label{en}
\ea 

\begin{figure}[hbt]
\includegraphics[scale=0.35]{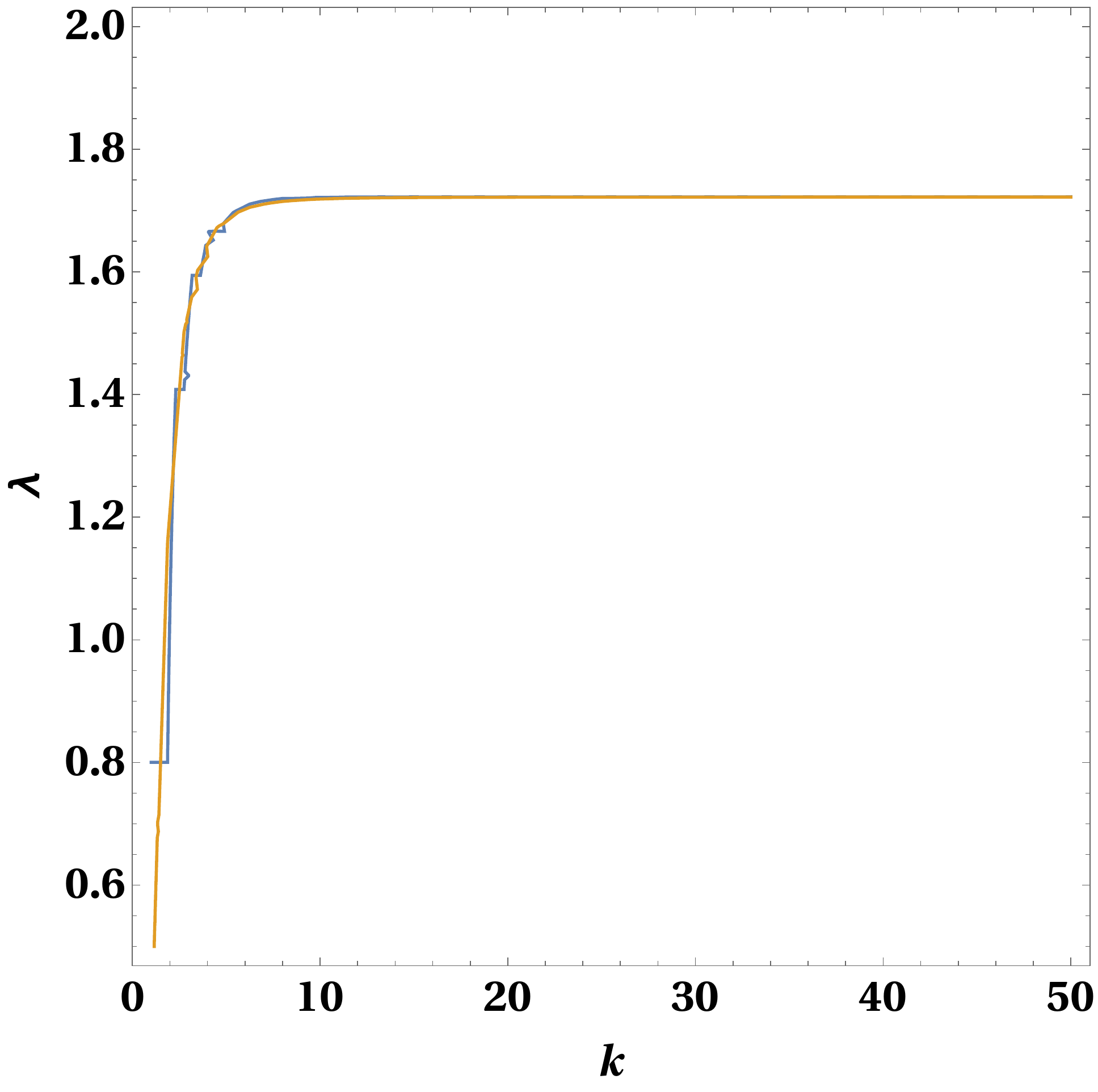}
\caption{\label{plot}} The blue curve shows the variation of $\lm$ as a function of $k$ obtained from eq.(\ref{lmk}). The yellow curve shows the plot of the function $\lm_0\exp{-\f{\alpha_0}{(k + k_0)^{\nu_0}}}$ of $k$, with $\lm_0=1.7220127,~ \alpha_0=27,~ k_0=1,~ \nu_0=4.$
\end{figure}

Now, a sum over $j$ on both sides of eq.(\ref{mpc}) leads to 
 the following consistency condition:
\ba
\sum_{j=1/2}^{k/2}(2j+1)e^{-\lm\j}=1\label{lmk}
\ea
Eq.(\ref{lmk}), in principle, should lead to a solution for $\lm$ as a function of $k$. To avoid the mathematical complication of finding this solution analytically, I plot\footnote{All the plots have been drawn by using Mathematica.} the function
\ba
\lm=\lm_0\exp{-\f{\alpha_0}{(k + k_0)^{\nu_0}}}, \label{lmsol}
\ea
where $\lm_0, \alpha_0, k_0$ and $\nu_0$ are some numbers. The plot (yellow coloured in fig.(\ref{plot})) fits to the curve obtained by plotting $\lm$  vs $k$ from eq.(\ref{lmk}), up to a `very good' approximation, for
\ba
\lm_0=1.7220127,~ \alpha_0=27,~ k_0=1,~ \nu_0=4.
\ea
I do not provide here a mathematical estimate of `how good' a fit it is. This is just an `optical' fit obtained by numerical experiments.

It is manifest from the fig. (\ref{plot}) that the curves almost merge together for $k\geq 2$. Further, for black holes with $A_c\gg\sO(G)$ one has $k\gg1$ as, one will see shortly that, $\g$ is a number of order unity. Henceforth, I shall consider eq.(\ref{lmsol}) as the functional dependence of $\lm$ on $k$.
\subsection{The BHAL} \label{bhal}
From eq.(\ref{en}), it follows that the entropy is given by the BHAL i.e.
\ba
S=\f{A_c}{4G},
\ea
if  the rescaling occurs as follows:
\ba
&&\s(k,\g)=\f{\lm(k)}{2\pi\g}=\f{\lm_0}{2\pi\g}\exp{-\f{\alpha_0}{(k+ k_0)^{\nu_0}}}.\label{sigf}
\ea
Recalling that $k:=A_c/4\pi\g G$, one can check that for all values of $\g$ within the range $0<\g\leq A_c/4\pi G$, $\s$ is positive definite which is needed for the positive definiteness of the rescaled area spectrum. This was property (i) enlisted at the end of section \ref{sec3}. Hence, I conclude that the statistical mechanical calculation with the horizon degrees of freedom in LQG leads to  the BHAL {\it without having to choose $\g$}, if the area spectrum of the black hole horizon is rescaled by $\s(k,\g)$ given by eq.(\ref{sigf}).

However, $\s$ needs to satisfy property (ii) as a requirement of physical consistency, as I shall explain in the next subsection. Importantly, I mention again and emphasize that property (ii) is independent of property (i).

\subsection{Determining $\g$ from the `fixed point'}
All standard QFTs are some effective field theories valid until some energy scale. Only the renormalized quantities are calculable and measurable. The bare values of those physical quantities can not be theoretically calculated. This is not unexpected because one does not have access to the most fundamental theory from which the corresponding QFT has come out to be an effective one. Taking quantum electrodynamics (QED) as an example, the bare electron charge is never measurable because one can not decouple the electron from its field.  However, if one {\it would have} known the most fundamental theory from which QED emerges effectively in some limit, then one could have expected to know, at least theoretically, the bare charge value of the electron. Added to this, the renormalized charge must have {\it asymptotically} approached that bare value in the high energy limit. 

In the present scenario, one has the `bare' area spectrum of an arbitrary 2-sphere and the rescaled (`renormalized') one on the black hole horizon. This is because one is now dealing with LQG which is one of the candidates of the fundamental theory of gravity. Hence, the `bare' quantities are expected to be known in this theory.
Therefore, it seems quite logical to demand that  $\lim_{k\to\infty}\s(k,\g)=1$. To mention again, the physics underlying this limit is the following. The coupling strength $1/k$ of the point like sources to the CS field strength decreases i.e. $1/k\to 0$. Hence, the area spectrum of the black hole should {\it asymptotically} approach the one of an arbitrary 2-sphere (the `bare' spectrum) in this limit. In fact, one can view this limit as the `fixed point' of the corresponding `renormalization group flow' on the horizon i.e. where the `beta function' corresponding to the running  gravitational constant on the horizon vanishes viz.
\ba
\f{d \tilde G(k)}{d(\ln k)}=0.\label{beta}
\ea 
From eq.(\ref{beta}), one can conclude that the `fixed point' is implied by the limit $k\to \infty$.

One should be aware of the fact that this is an asymptotic limit and $\s$ is never exactly unity since $k$ is never exactly infinity. Putting $k=\infty$ in eq.(\ref{bc}) will give $F=0$ as the field equation on the horizon indicating that the sources have completely decoupled, which does not hold any meaning. This is analogous to the fact that if the bare charge of the electron {\it were} known, the renormalized charge would have only asymptotically approached that value in the high energy limit. However, it would have never exactly matched the bare value of the charge because that would have meant the electron has  decoupled from its own field.

Now, using eq.(\ref{G}) and eq.(\ref{sigf}) it is trivial to check that in this limit i.e. at the fixed point, $\tilde G$ asymptotically approaches $\g G$ only for a particular value of $\gamma$:  
\ba
&&\tilde G|_{\text{fixed point}}=\g G\nn\\
&\implies&\lim_{k\to\infty}\s(k,\g)=1\nn\\
&\implies& \g=\lm_0/2\pi.~~
\ea
 It may be noted that this is the exact value of $\g$ that had to be {\it chosen} to obtain the BHAL in the usual practice \cite{sigma}.  However, the difference is that, as one can see, now $\g$ is {\it determined} by a  physical consistency requirement associated with the running gravitational constant on the horizon rather than being chosen to match a result.

{\it Few comments:} I shall make a digression here to offer some comments in relation to the available literature. It is very important to note that the present scenario is completely different from the one that was proposed in \cite{jac}. The renormalization of gravitational constant proposed in \cite{jac} is related to the renormalization of the fundamental degrees of freedom of LQG theory resulting in  the general relativity emerging in the effective field theory limit. In this scenario there is a possibility that the gravitational constant can depend on the area of the black hole which creates the following problem \cite{daniel}: what is the gravitational constant for a spacetime with more than one black hole? On the contrary, in the present scenario, I have proposed a `renormalization' effect taking place only on the horizon due to the associated quantum theory.  Since this effect is localized on the horizon, no problem arises in the presence of more than one black holes. 

\section{Conclusion}
Whatever I have discussed here is purely based on heuristic arguments that rely on some observations of the field theoretic structure that effectively describes the black hole horizon degrees of freedom in LQG and some analogies. This by no means is anything mathematically rigorous. However, having the knowledge of the full dynamics of quantum black holes in LQG yet out of reach, such a possibility of a `renormalized' gravitational constant governing the microscopic  physics on the horizon and giving rise to the BHAL irrespective of the value of $\g$, can not be ruled out completely. Also, I emphasize that the value of $\g$ reported here has been {\it determined} by studying the asymptotic limit, of the `renormalized' gravitational constant on the horizon, in which it approaches the   
`bare' value. The limit can viewed as the fixed point for the `renormalization group flow' on the horizon i.e. the beta function corresponding to the running gravitational constant on the horizon vanishes in this limit.  Unlike the usual practice this is {\it not a choice} of $\g$ to match the entropy with the BHAL. I hope this work may give a possible hint towards a more fundamental calculation of black hole entropy from LQG involving the underlying dynamics of quantum black hole horizons leading to the BHAL irrespective of the choice of $\g$.

\vspace{0.1cm}
~\\
{\bf Acknowledgments:} I am grateful to Benito Juarez Aubry and Tim Koslowski for carefully going through the manuscript and offering critical comments that have helped me in improving the same to a large extent. Most of the work was supported by DGAPA postdoctoral fellowship of UNAM, Mexico.

\end{document}